\documentstyle[12pt,epsfig]{article}
\def\be{\begin{equation}} \def\ee{\end{equation}} \def\bea{\begin{eqnarray}}
\def\eea{\end{eqnarray}} \def\nnb{\nonumber}

\begin{document}
%

\hfill{October 4, 2007}

\begin{center}
\vskip 6mm 
{\Large\bf 
Low energy proton-proton scattering  \\
in effective field theory}
\vskip 6mm 
{\large 
Shung-ichi Ando\footnote{mailto:sando@color.skku.ac.kr}, 
Jae Won Shin, Chang Ho Hyun, and Seung Woo Hong}
\vskip 6mm 
{\large \it Department of Physics, 
Sungkyunkwan University,
Suwon 440-746, Korea}

\end{center}

\vskip 5mm

Low energy proton-proton scattering 
is studied in pionless effective field theory. 
Employing 
the dimensional regularization and $\overline{\mbox{\rm MS}}$ and
power divergence subtraction schemes
for loop calculation, 
we calculate the scattering amplitude in $^1S_0$ channel 
up to next-to-next-to leading order 
and fix
low-energy constants that appear in the amplitude
by effective range parameters. 
We study regularization scheme and scale dependence 
in separation of Coulomb interaction from 
the scattering length and effective range
for the $S$-wave proton-proton scattering. 

\vskip 3mm \noindent
{PACS(s): 11.10.Gh, 13.75.Cs.}

\newpage \noindent 
{\bf 1. Introduction} 

Effective field theories (EFTs),
which provide us a systematic perturbative scheme and 
a model-independent calculation method, 
have become a popular method to 
study hadronic reactions 
with and without external probes at low and intermediate energies.
(See, {\it e.g.},  
Refs.~\cite{betal-01,bvk-arnps02,kp-arnps04,e-ppnp06,b-07} for reviews.)   
At very low energies, 
the Coulomb interaction becomes essential
for the study of reactions involving charged particles.
The first consideration of the Coulomb interaction 
in a pionless EFT was done by Kong and Ravndal (KR)
for low energy $S$-wave proton-proton ($pp$) 
scattering~\cite{kr-plb99,kr-npa00}.
They calculated the $pp$ scattering amplitude
up to next-to leading order (NLO).
For loop calculations, 
they employed
dimensional regularization 
with minimum subtraction (MS) scheme 
and so called power divergence subtraction (PDS) scheme
suggested by Kaplan, Savage and Wise~\cite{ksw-plb98,ksw-npb98}.
Then KR estimated a scattering length $a(\mu)$ 
for the $pp$ scattering 
after separating off
the Coulomb correction 
where $\mu$ is the scale for dimensional regularization.
The leading order (LO) result of $a(\mu)$ 
was almost infinite at $\mu=m_\pi$
where $m_\pi$ is the pion mass~\cite{kr-plb99}.
In addition, the LO $a(\mu)$ was highly dependent on the value of $\mu$.
Including the NLO correction, they obtained
$a(\mu=m_\pi)=-29.9$ fm~\cite{kr-npa00}
which is comparable to the value of 
the scattering length $a_{np}$ in the $np$ channel,
$a_{np}=-23.748\pm 0.009$ 
fm~\footnote{See, e.g., Table VIII in Ref.~\cite{av18}.}.

The value of $a(\mu)$ 
deduced after separating the Coulomb and strong interactions 
is particularly important in the study of isospin breaking effects 
in $S$-wave $NN$ interaction~\cite{sat-pr89,mns-pr90}. 
The accurate value of $a_{np}$ is well known as 
quoted above, 
while the values of the scattering length in 
the $nn$ channel ($a_{nn}$) and in the $pp$ channel ($a_{pp}$)
still have considerable uncertainties.

There exists no direct $nn$ scattering experiment
because of the lack of {\it free} neutron target. 
The values of $a_{nn}$ have been 
deduced from the experimental data of $\pi^-d\to nn\gamma$ and 
$nd\to nnp$ reactions. 
Recent publications suggest 
$a_{nn} = -18.50\pm0.05(stat.)\pm0.44(syst.)\pm0.30(th.)$ fm 
from the $\pi^-d\to nn\gamma$ process~\cite{hetal-plb98} 
and $a_{nn}= -18.7\pm0.6$ fm~\cite{getal-prl99}, 
$-16.06\pm 0.35$ fm~\cite{hetal-prl00} and 
$-16.5\pm 0.9$ fm~\cite{wetal-prc06} from 
the $nd\to nnp$ process.
As seen, the values of $a_{nn}$ have significant errors 
compared to that of $a_{np}$, and 
the center values do not seem to converge yet.\footnote{
Recently, there were proposals
to determine the value of $a_{nn}$
more precisely by 
employing a formalism of EFT,
from the $\pi^-d\to nn\gamma$ reaction\cite{gp-prc06} 
and neutron-neutron fusion, $nn\to de^-\bar{\nu}_e$\cite{ak-plb06}.
}

For the $pp$ channel, 
a very accurate value of the scattering length 
$a_C=-7.828\pm 0.008$ fm~\cite{aetal-prc84} 
and $a_C= - 7.8149\pm 0.0029$ fm~\cite{m-prc01} are available
from the low energy $pp$ scattering data. 
It contains however 
contributions from both strong and electromagnetic 
interactions, and thus we need 
to disentangle the strong interaction 
from the electromagnetic interaction.
It was pointed out in potential model calculations
that there is a considerable model dependence in deducing the
value of the strong scattering length $a_{pp}$ 
from $a_C$~\cite{aetal-prc84,app}.
Some literature shows  
$a_{pp}= -17.1\pm 0.2$ fm~\cite{aetal-prc84},  
while a heavy-baryon chiral perturbation theory 
results in $a_{pp}=-17.51\sim -16.96$ fm~\cite{wme-npa01}
with uncertainties slightly larger than those from the 
potential models.

In this work, we employ the pionless EFT~\cite{crs-npa99} 
including the Coulomb interaction between two protons~\cite{kr-plb99,kr-npa00}
and calculate the $pp$ scattering amplitude
with the strong $NN$ interactions up to next-to-next-to leading order
(NNLO).
Our main motivation of this study is 
to see how the value of strong scattering length $a(\mu=m_\pi)=-29.9$ fm
obtained by KR from NLO calculations
may be improved by the inclusion of a higher order correction. 
We find that the NNLO corrections turn out to be quite small 
but there is a considerable dependence 
of the scattering length $a(\mu)$ 
on the renormalization schemes and the scale parameter $\mu$. 

This paper is organized as follows.
In Sec. 2 we briefly review the effective range formalism 
for the $pp$ scattering.
In Sec. 3 the pionless strong effective Lagrangian up 
to NNLO is introduced. 
In Sec. 4 we calculate the $S$-wave $pp$ scattering amplitude
up to NNLO. 
In Sec. 5, we discuss regularization method and renormalization schemes
employed in this work.
We renormalize low energy constants (LECs) that appear
in the strong $NN$ interaction up to NNLO by effective range parameters
employing MS-bar ($\overline{\mbox{\rm MS}}$) and PDS schemes
and obtain numerical results for the strong scattering length $a(\mu)$
and strong effective range $r(\mu)$. 
Discussion and conclusions are given in Sec. 6.
In Appendix A we show detailed expressions of the amplitudes in NNLO.
Detailed calculations of the loop functions
employing the dimensional regularization and 
$\overline{\mbox{\rm MS}}$ 
and PDS schemes are given in Appendix B.

\vskip 3mm \noindent 
{\bf 2. Proton-proton scattering in effective range theory} 

The amplitude of the $pp$ scattering 
can be decomposed 
as~\cite{QCT75}
\bea
T = T_C + T_{SC}\, ,
\eea
where $T_C$ is the pure Coulomb part and 
$T_{SC}$ is the ``modified'' strong amplitude 
whose $S$-wave channel we calculate up to NNLO
in pionless EFT below.

The incoming and outgoing scattering states 
$|\Psi_{\vec{p}}^{(\pm)}\rangle$ with the potential
$\hat{V}=\hat{V}_C+\hat{V}_S$ where
$\hat{V}_C$ and $\hat{V}_S$ are the Coulomb and strong potentials,
respectively, 
are represented 
in terms of the Coulomb states $|\psi_{\vec{p}}^{(\pm)}\rangle$ as
\bea
|\Psi_{\vec{p}}^{(\pm)}\rangle = \sum_{n=0}^\infty 
(\hat{G}_C^{(\pm)}\hat{V}_S)^n |\psi_{\vec{p}}^{(\pm)}\rangle \, ,
\eea
where $\hat{G}_C^{(\pm)}$ is the incoming and outgoing Green's function 
\bea
\hat{G}_C^{(\pm)}(E) = \frac{1}{E-\hat{H}_0-\hat{V}_C\pm i\epsilon} \, .
\label{eq;GC}
\eea
Here
$\hat{H}_0=\hat{p}^2/M$
is the free Hamiltonian of two protons
and
$V_C= e^2/(4\pi r)$ is the repulsive Coulomb potential.
The Coulomb state $|\psi_{\vec{p}}^{(\pm)}\rangle$ is obtained by
solving the Schr\"{o}dinger equation 
$(\hat{H}-E)|\psi^{(\pm)}_{\vec{p}}\rangle =0$ 
with $\hat{H}=\hat{H}_0+\hat{V}_C$ and thus one has
\bea
|\psi_{\vec{p}}^{(\pm)}\rangle = \left[1+\hat{G}_C^{(\pm)}\hat{V}_C\right]
|\vec{p}\rangle\, ,
\eea 
where $|\vec{p}\rangle$ is the free wave state.
The normalization of $|\psi_{\vec{p}}^{(\pm)}\rangle$ is 
such that 
$\langle \psi_{\vec{p}}^{(\pm)}|\psi_{\vec{q}}^{(\pm)}\rangle 
= (2\pi)^3 \delta^{(3)}(\vec{p}-\vec{q})$.
The amplitude $T_{SC}$ 
is thus obtained by
\bea
T_{SC}(\vec{p}',\vec{p}) &=& 
\sum_{n=0}^\infty \langle \psi_{\vec{p}'}^{(-)}|
\hat{V}_S (\hat{G}_C^{(+)}\hat{V}_S)^n|\psi_{\vec{p}}^{(+)}\rangle\, .
\label{eq;TSC}
\eea
For $l=0$ state one has the amplitude
\bea
T_{SC}^{l=0} = - \frac{4\pi}{M}\frac{e^{2i\sigma_0}}{p\cot\delta_0-ip}\, ,
\eea
where $\sigma_l$ is the Coulomb phase shift
$\sigma_l= \arg \Gamma(1+l+i\eta)$
with $\eta = \alpha M/(2p)$. 
In the effective range expansion with the Coulomb interaction,
the modified strong phase shift $\delta_l$ for $l=0$
in low energy $pp$ scattering is represented 
by effective range parameters~\cite{b-pr49}:
\bea
C_\eta^2\, p\cot\delta_0 + \alpha M h(\eta)
= -\frac{1}{a_C} + \frac12 r_0 p^2 
- Pr_0^3 p^4 + \cdots\, ,
\label{eq;ERP}
\eea
where $C_\eta^2= 2\pi \eta/(e^{2\pi \eta}-1)$ 
and
\bea 
h(\eta)={\rm Re}\, \psi(i\eta)-\ln\eta\, .
\label{eq;heta}
\eea
$\psi$-function is the logarithmic 
derivative of the Gamma function and  
${\rm Re}\, \psi(i\eta) = 
\eta^2\sum_{\nu=1}^\infty \frac{1}{\nu(\nu^2+\eta^2)}
-C_E$; 
$C_E$ is the Euler's constant, $C_E= 0.577215\cdots$.
Effective range parameters
$a_C$, $r_0$, $P$ are 
modified scattering length, effective range, effective volume,
respectively. 

\vskip 3mm \noindent 
{\bf 3. Effective Lagrangian} 
\vskip 2mm

Pionless effective Lagrangian 
for strong $S$-wave $NN$ interaction up to
NNLO reads~\cite{crs-npa99,fms-npa00}
\bea
{\cal L} &=& N^\dagger \left( 
iD_0 + \frac{\vec{D}^2}{2m_N}\right) N
-C_0 \left[N^T P_a^{({}^1S_0)} N\right]^\dagger N^T P_a^{({}^1S_0)}N
\nnb \\ && + \frac12 C_2 
\left[N^T P_a^{({}^1S_0)} \stackrel{\leftrightarrow}{D}^2 
N\right]^\dagger N^T P_a^{({}^1S_0)}N
+ h.c.
\nnb \\ &&
- \frac12 C_4
\left(N^TP_a^{({}^1S_0)}\stackrel{\leftrightarrow}{D}^2N\right)^\dagger
N^TP_a^{({}^1S_0)}\stackrel{\leftrightarrow}{D}^2N
\nnb \\ &&
- \frac14 \tilde{C}_4
\left[
\left(N^TP_a^{({}^1S_0)}\stackrel{\leftrightarrow}{D}^4N\right)^\dagger
N^TP_a^{({}^1S_0)}N 
+ h.c.
\right]   \, ,
\label{eq;L}
\eea
where $D_\mu$ is the covariant derivative, 
$\stackrel{\leftrightarrow}{D} = \frac12 (\stackrel{\rightarrow}{D}
-\stackrel{\leftarrow}{D})$,
and $P_a^{({}^1S_0)}$ is a projection operator for the two-nucleon
${}^1S_0$ states, $P_a^{({}^1S_0)}=\frac{1}{\sqrt{8}}\sigma_2\tau_2\tau_a$.
Note that we retain two low energy constants, 
$C_4$ and $\tilde{C}_4$, in NNLO.

The strong $NN$ potential is
expanded in terms of small momentum as
\bea
\hat{V}_S = \hat{V}_0 + \hat{V}_2 + \hat{V}_4 + \cdots\, ,
\label{eq;VS}
\eea
where $\hat{V}_0$, $\hat{V}_2$, $\hat{V}_4$ are LO, NLO, NNLO potential,
respectively, and the matrix elements of them are obtained from the 
Lagrangian in Eq.~(\ref{eq;L}) as 
\bea
\langle \vec{q}|\hat{V}_0|\vec{k}\rangle &=& C_0\, ,
\label{eq;V0}
\\
\langle \vec{q}|\hat{V}_2|\vec{k}\rangle &=& 
\frac12 C_2 (\vec{q}^2+\vec{k}^2)\, ,
\\
\langle \vec{q}|\hat{V}_4|\vec{k}\rangle &=& 
\frac12 C_4 \vec{q}^2\vec{k}^2 
+ \frac14 \tilde{C}_4 (\vec{q}^4 
+ \vec{k}^4) \, ,
\label{eq;V4}
\eea
where $|\vec{q}\rangle$ and $|\vec{k}\rangle$ are the intermediate 
free two-nucleon outgoing and incoming states, respectively:
2$\vec{q}$ and 2$\vec{k}$ are the relative momenta for 
the two protons.

In this work we employ the standard counting rules 
of the strong $NN$ interaction 
with the PDS scheme 
in Refs.~\cite{kr-npa00,ksw-plb98}.
(We will discuss the PDS scheme in detail later.)
For the strong potential, 
the LO term $C_0$ is counted as $Q^{-1}$ order,
where $Q$ denotes the small expansion parameter, 
and is summed up to an infinite order.
The NLO ($C_2$) and NNLO ($C_4$, $\tilde{C}_4$) terms 
are counted
as $Q^2$ and $Q^4$, respectively, and expanded perturbatively.\footnote{
Note that 
by changing the LECs $C_4$ and $\tilde{C}_4$ in another linear 
combination, e.g.,
$C_4=C_4'+\tilde{C}_4'$ and $\tilde{C}_4=C_4'-\tilde{C}_4'$,
one can easily see that the term proportional to $\tilde{C}_4'$ in
Eq.~(\ref{eq;V4}) vanishes when $|\vec{q}|=|\vec{k}|$.
The $\tilde{C}_4'$ term, so called off-shell term, is 
redundant and vanishes when the external legs of the potential
go on mass-shell.
\label{footnote;C4tildeC4} 
}
We treat the Coulomb interaction 
non-perturbatively 
using the Green's function 
$G_C^{(\pm)}$ in Eq.~(\ref{eq;GC}).
We do not include higher order QED corrections such as 
the vacuum polarization effects reported in Refs.~\cite{vp}.

\vskip 2mm \noindent 
{\bf 4. Amplitudes} 
\vskip 1mm

The amplitude $T_{SC}^{l=0}$ for the $S$-wave $pp$ scattering 
can be written as 
\bea
T_{SC}^{l=0} = T_{SC}^{(0)} + T_{SC}^{(2)} + T_{SC}^{(4)} + \cdots\, ,
\eea
where $T_{SC}^{(0)}$, $T_{SC}^{(2)}$, $T_{SC}^{(4)}$ 
are LO, NLO, NNLO amplitudes, respectively. 
By inserting the strong LO potential $\hat{V}_0$ 
in Eq.~(\ref{eq;V0}) 
into the amplitude $T_{SC}$ 
in Eq.~(\ref{eq;TSC}),
we obtain the LO amplitude $T_{SC}^{(0)}$ in terms of loop 
functions $\psi_0$ and $J_0$:
\bea
T_{SC}^{(0)} &=& \sum_{n=0}^\infty \langle \psi_{\vec{p}'}^{(-)}|
\hat{V}_0 (\hat{G}_C^{(+)}\hat{V}_0)^n|\psi_{\vec{p}}^{(+)}\rangle
= \frac{C_0
\psi_0^2(p)
}{1-C_0J_0(p)}\, ,
\eea
where
\bea
\psi_0(p) &=& 
\int \frac{d^3\vec{k}}{(2\pi)^3}\psi_{\vec{p}}^{(+)}(\vec{k})
= \int \frac{d^3\vec{k}}{(2\pi)^3}\psi_{\vec{p}}^{(-)*}(\vec{k}) \, ,
\label{eq;psi0}
\\
J_0(p) &=&  
\int \frac{d^3\vec{k}'}{(2\pi)^3}
\frac{d^3\vec{q}}{(2\pi)^3} 
\langle \vec{q}|\hat{G}_C^{(+)}|\vec{k}'\rangle \,  .
\label{eq;J0}
\eea
Detailed calculations for the functions
$\psi_0$ and $J_0$ are given in Appendix B.
$T_{SC}^{(0)}$ is summation of 
the LO strong potential $\hat{V}_0$,
that is, the $C_0$ terms summed up to the infinite order.

\begin{figure}[h]
\begin{center}
\epsfig{file=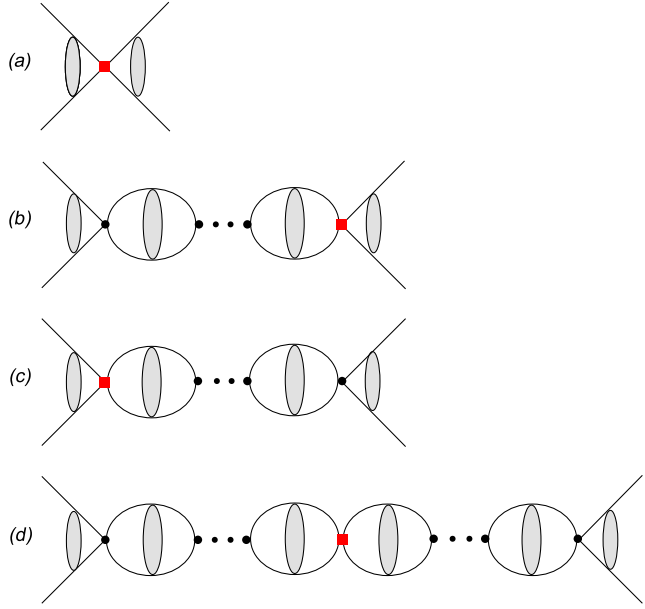, width=8.5cm}
\caption{
NLO diagrams for the $S$-wave $pp$ scattering.
Gray blobs denote the two-proton Coulomb Green's function $G_C^{(+)}$, 
and two nucleon contact vertices denote the strong potential: 
the (black) circle and
the (red) square represent LO ($C_0$) and NLO ($C_2$)
vertices, respectively.
Small double dots stand for the summation of $C_0$ terms
up to the infinite order.}
\label{fig;nlo-1}
\end{center}
\end{figure}
At NLO we have four diagrams shown in 
Fig.~\ref{fig;nlo-1}.\footnote{
Figures were prepared using the program JaxoDraw~\cite{jaxodraw}
provided by L. Theussl.
}
They are proportional to $C_2$ coming from $V_2$, 
whereas the $C_0$ terms are summed up to the infinite order.
The NLO amplitude is written in terms of the loop functions
$\psi_0$, $\psi_2$, $J_0$ and $J_2$ as
\bea
T_{SC}^{(2,a-d)} &=& 
\frac{C_2\psi_0}{(1-C_0J_0)^2}
\left[\psi_2 + C_0(\psi_0J_2 - \psi_2J_0)
\right]\, ,
\eea
with
\bea
\psi_2(p) &=&  
\int\frac{d^3\vec{k}}{(2\pi)^3}\vec{k}^2 \psi_{\vec{p}}^{(+)}(\vec{k})
= \int\frac{d^3\vec{k}}{(2\pi)^3}\vec{k}^2 \psi_{\vec{p}}^{(-)*}(\vec{k}) \, , 
\label{eq;psi2}
\\
J_2(p) &=& 
\int \frac{d^3\vec{q}}{(2\pi)^3}
\frac{d^3\vec{q}'}{(2\pi)^3}
\vec{q}^{'2}\langle \vec{q}'|\hat{G}_C^{(+)}|\vec{q}\rangle
=\int \frac{d^3\vec{q}}{(2\pi)^3}
\frac{d^3\vec{q}'}{(2\pi)^3}
\langle \vec{q}'|\hat{G}_C^{(+)}|\vec{q}\rangle
\vec{q}^{2} \, .
\label{eq;J2}
\eea
Details for $\psi_2$ and $J_2$ are given in Appendix B.
The NLO amplitude $T_{SC}^{(2)}$ consists of 
one $C_2$ and a summation of the $C_0$ terms up to the infinite order.
These LO and NLO amplitudes 
have already been obtained by KR
in Ref.~\cite{kr-npa00}. 

\begin{figure}[h]
\begin{center}
\epsfig{file=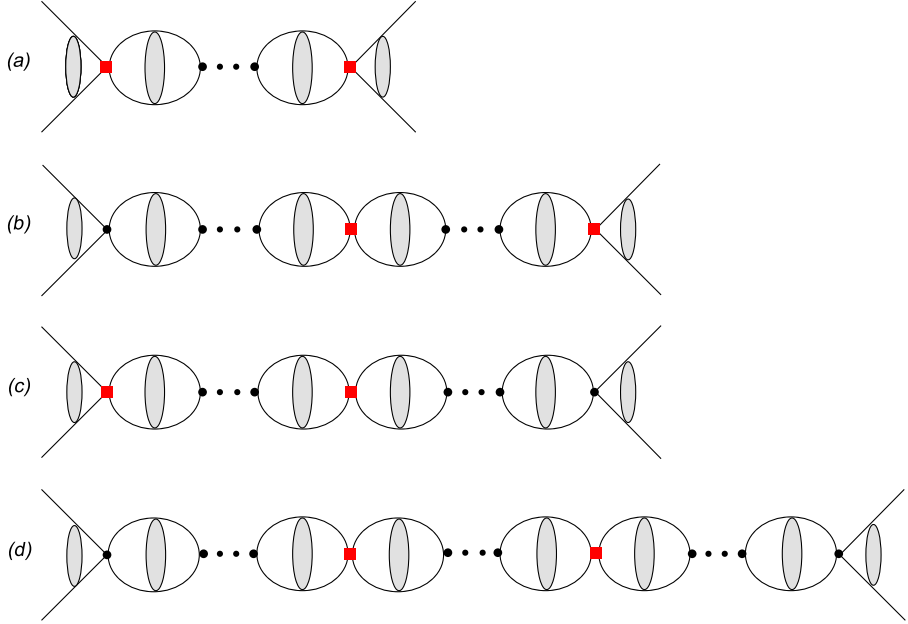, width=12cm}
\caption{Set 1 of NNLO diagrams. 
See the caption of Fig.~\ref{fig;nlo-1} for details.}
\label{fig;nnlo-1}
\end{center}
\end{figure}

\begin{figure}[h]
\begin{center}
\epsfig{file=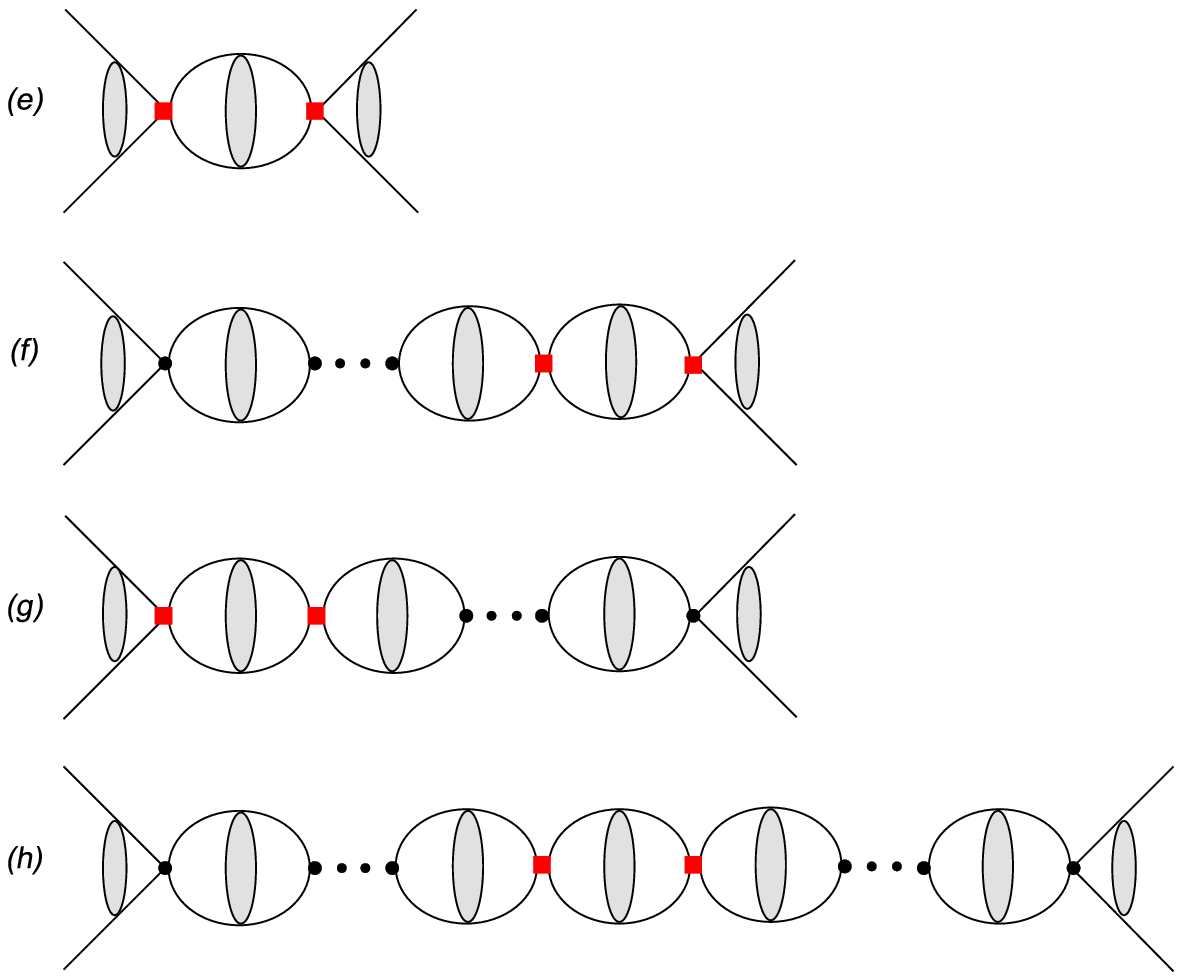, width=11cm}
\caption{
Set 2 of NNLO diagrams.
See the caption of Fig.~\ref{fig;nlo-1} for details.}
\label{fig;nnlo-2}
\end{center}
\end{figure}

\begin{figure}[h]
\begin{center}
\epsfig{file=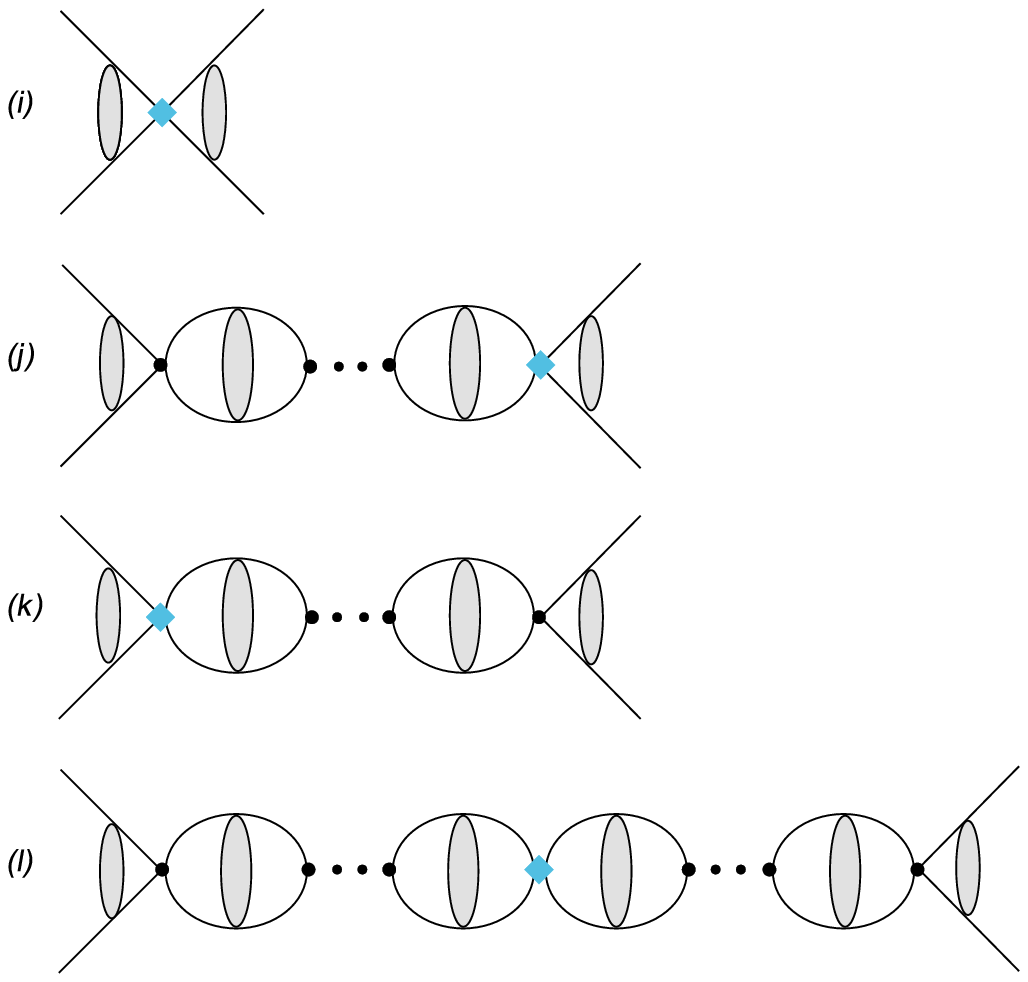, width=9cm}
\caption{
Set 3 of NNLO diagrams.
Two-proton contact vertices represented by
(blue) diamonds denote
strong NNLO potential $V_4$. 
See the caption of Fig.~\ref{fig;nlo-1} 
for details.}
\label{fig;nnlo-3}
\end{center}
\end{figure}

At NNLO we have three sets of diagrams 
shown in Figs.~\ref{fig;nnlo-1}, \ref{fig;nnlo-2},
and \ref{fig;nnlo-3}.
From the first and second sets of diagrams shown 
in Figs.~\ref{fig;nnlo-1} and \ref{fig;nnlo-2}, respectively,
we see two NLO corrections to the amplitude 
and thus the NNLO amplitudes 
obtained from the first and second sets of diagrams 
in Figs.~\ref{fig;nnlo-1} and \ref{fig;nnlo-2}
are proportional to $C_2^2$. 
The NNLO amplitudes 
corresponding to the diagrams in Fig.~\ref{fig;nnlo-1} 
can be written 
in terms of the functions $\psi_0$, $\psi_2$, $J_0$ and $J_2$,
whereas to express the amplitudes 
for the diagrams 
in Fig.~\ref{fig;nnlo-2}  we need a new function $J_{22}$
given below.
In the third set of diagrams shown in Fig.~\ref{fig;nnlo-3},
we have one NNLO correction to the amplitude
and the NNLO amplitudes 
for the diagrams in Fig.~\ref{fig;nnlo-3}
are proportional to $C_4$ or $\tilde{C}_4$.
Explicit expressions of the NNLO amplitude from each of the diagrams 
are given 
in terms of $\psi_i$ with $i=0,2,4$ and $J_j$ with $j=0,2,22,4$ 
in Appendix A.
 
Summing up the amplitudes obtained from the diagrams (a) to (h) 
in Figs.~\ref{fig;nnlo-1} and \ref{fig;nnlo-2} we have
\bea
T_{SC}^{(4,a-h)} &=& 
\frac{C_2^2}{4(1-C_0J_0)^3}
\left\{
\psi_0^2J_{22}(1-C_0J_0)
+ \psi_2^2J_0(1-C_0J_0)^2
\right.
\nnb \\ && 
\left.
+2\psi_0\psi_2J_2(1-C_0^2J_0^2)
+ \psi_0^2 J_2(C_0J_2) (3+C_0J_0)
\right\} \, ,
\eea
where 
\bea
J_{22} &=& 
\int \frac{d^3\vec{q}}{(2\pi)^3} 
\frac{d^3\vec{q}'}{(2\pi)^3} 
\vec{q}'^2 
\langle\vec{q}'|\hat{G}_C^{(+)}|\vec{q}\rangle 
\vec{q}^2\, ,
\label{eq;J22}
\eea
whose details are given in Appendix B.

Summing up the amplitudes for the diagrams (i) to (l) 
in Fig.~\ref{fig;nnlo-3} gives us
\bea
T_{SC}^{(4,i-l)} &=& 
\frac12 \frac{C_4}{(1-C_0J_0)^2}\left[
\psi_2^2(1-C_0J_0)^2
+ 2\psi_0\psi_2C_0J_2(1-C_0J_0)
+ \psi_0^2C_0^2J_2^2
\right] 
\nnb \\ &&
+ \frac12 \frac{\tilde{C}_4}{(1-C_0J_0)^2}
\left[ \psi_4 + C_0(\psi_0J_4-\psi_4J_0) \right]\psi_0\, ,
\eea
where 
\bea
\psi_4 &=& 
\int\frac{d^3\vec{k}}{(2\pi)^3}
\psi_{\vec{p}}^{(-)*}(\vec{k})\vec{k}^4
=\int\frac{d^3\vec{k}}{(2\pi)^3}
\vec{k}^4
\psi_{\vec{p}}^{(+)}(\vec{k})\, ,
\label{eq;psi4}
\\
J_4 &=& 
\int\frac{d^3\vec{q}}{(2\pi)^3} 
\frac{d^3\vec{q}'}{(2\pi)^3} 
\vec{q}'^4
\langle \vec{q}'|\hat{G}_C^{(+)}|\vec{q}\rangle
=\int\frac{d^3\vec{q}}{(2\pi)^3} 
\frac{d^3\vec{q}'}{(2\pi)^3} 
\langle \vec{q}'|\hat{G}_C^{(+)}|\vec{q}\rangle
\vec{q}^4 \, .
\label{eq;J4}
\eea
Calculations of $\psi_4$ and $J_4$ are given in Appendix B.

\vskip 3mm \noindent
{\bf 5. Regularization method and renormalization schemes}

In the calculation of the loop functions $J_0$,
$J_2$, $J_{22}$ and $J_4$ in Eqs.~(\ref{eq;J0}),
(\ref{eq;J2}), (\ref{eq;J22}), (\ref{eq;J4}),
we encounter infinities 
and employ the dimensional regularization.
We also employ the PDS scheme,
suggested by Kaplan, Savage and Wise~\cite{ksw-plb98,ksw-npb98}, 
in which one subtracts the poles in $d=3$ as well as those in 
$d=4$ space-time dimensions so that one obtains an expected 
perturbation series in the expansion of the $NN$ potential
in Eq.~(\ref{eq;VS}) with a given scale $\mu$ of the theory.
We may check the convergence radius, e.g., 
for the $C_2$ term (relative to the $C_0$ term) 
in Eq.~(\ref{eq;VS}) and have
$\Lambda_{20}(\mu) \equiv \sqrt{C_0(\mu)/C_2(\mu)}= 147$ 
(30.6) MeV with (without) the PDS terms at $\mu=m_\pi$.
Thus a formal convergence of the perturbative series 
of the $NN$ potential in Eq.~(\ref{eq;VS})
is improved thanks to the PDS term,
and the theory would be valid up to 
$p\sim \Lambda_{20} \simeq 140$ MeV,
which is the large scale we assumed in the pionless theory. 

The loop functions can be decomposed into 
a finite term and an infinite one, e.g.
$J_0=J_0^{fin}+J_0^{div}$ with 
$J_0^{fin}=-\frac{\alpha M^2}{4\pi}H(\eta)$ 
(the definition of the $H(\eta)$ function is given in Appendix B)
and 
\bea
J_0^{div} &=& 
-\frac{M}{4\pi}\mu 
+ \frac{\alpha M^2}{8\pi}\left[
\frac{1}{\epsilon}
-3 C_E 
+ 2
+\ln\left(\frac{\pi\mu^2}{\alpha^2M^2}\right)
\right] \, ,
\label{eq;J0div}
\eea
where $J_0^{div}$ is calculated by 
the dimensional regularization in $d=4-2\epsilon$ dimensions 
and the PDS scheme.
The first term proportional to the scale $\mu$ in the r.h.s. 
of Eq.~(\ref{eq;J0div}) is the PDS term
and $C_E$ is the Euler's constant mentioned 
earlier.
The scattering amplitudes should be 
identical after renormalization 
even if another renormalization scheme such as 
off-shell momentum subtraction scheme discussed 
in Refs.~\cite{fms-npa00,g-98} is employed.
However, $a(\mu)$ and $r(\mu)$ 
do depend on the renormalization schemes
along with the value of the renormalization scale $\mu$.
So, to be consistent with KR,
we calculate all the loop functions $J_i$ 
with $i=0, 2, 22, 4$ and the wavefunctions $\psi_j$
with $j=0, 2, 4$
by using the dimensional regularization and 
the PDS scheme in Appendix B.

The $S$-wave $pp$ scattering amplitude in terms of
the effective range parameters is given by 
\bea
T_{SC}^{l=0} &=& 
- \frac{4\pi}{M} \frac{C_\eta^2e^{2i\sigma_0}}{
-\alpha MH(\eta) -\frac{1}{a_C} + \frac12 r_0p^2-Pr_0^3p^4+\cdots}\, ,
\eea
and thus one has
\bea
\lefteqn{
-\frac{1}{a_C} + \frac12 r_0p^2 -Pr_0^3p^4 + \cdots
=
\alpha MH(\eta)
-\frac{4\pi}{M}C_\eta^2e^{2i\sigma_0}\frac{1}{T_{SC}^{l=0}}
}
\nnb \\ &=& 
\alpha MH(\eta)
-\frac{4\pi}{M} 
\frac{\psi_0^2}{T_{SC}^{(0)}} 
\left[ 1
- \frac{T_{SC}^{(2)}}{T_{SC}^{(0)}}
- \frac{T_{SC}^{(4)}}{T_{SC}^{(0)}}
+ \left( \frac{T_{SC}^{(2)}}{T_{SC}^{(0)}}\right)^2
+\cdots
\right] \, .
\label{eq;renormERP}
\eea
Comparing the coefficients of the terms proportional to 
$p^0$, $p^2$ and $p^4$ in both sides of Eq.~(\ref{eq;renormERP}), 
we have 
\bea
-\frac{1}{a_C} &=& -\frac{4\pi}{M}\left\{
\frac{1}{C_0} -J_0^{div}
+\frac{C_2}{C_0^2}\left[
\alpha M\mu + \frac12(\alpha M)^2 
+C_0\frac{\pi M}{48}(\alpha M)^2\mu\right]
\right. \nnb \\ && \left.
-\left(
\frac12 \frac{C_4}{C_0^2} 
-\frac{C_2^2}{C_0^3}\right) 
(\alpha M)^2\mu^2\right\} 
+{\cal O}(\alpha^3)\, ,
\label{eq;aC}
\\
+\frac12 r_0 &=&
\frac{4\pi}{M}\left[
\frac{C_2}{C_0^2} 
-2 \left(\frac12 \frac{C_4}{C_0^2}
+\frac13 \frac{\tilde{C}_4}{C_0^2}
- \frac{C_2^2}{C_0^3}\right)(\alpha M)\mu
\right] + {\cal O}(\alpha^2)\, ,
\label{eq;r0}
\\
-Pr_0^3 &=& \frac{4\pi}{M}\left(
\frac12\frac{C_4}{C_0^2}
+\frac12\frac{\tilde{C}_4}{C_0^2}
-\frac{C_2^2}{C_0^3}\right)\, ,
\label{eq;P}
\eea
where we have expanded the r.h.s. of 
Eqs.~(\ref{eq;aC}) and (\ref{eq;r0})  
in the order of the fine structure constant $\alpha$
and neglected the $\alpha^3$ ($\alpha^2$) and higher order terms 
in Eq.~(\ref{eq;aC}) (Eq.(\ref{eq;r0})).
With three effective range parameters, we cannot determine the four
LECs uniquely.
There are some arguments which can constrain the values of $C_4$ and
$\tilde{C}_4$.
The $C_4$ contribution in Eq.~(\ref{eq;aC}) is of 
the order of $\mu^2$,
and thus the first $\tilde{C}_4$ contribution term 
is of the lower order of $\mu$ than the $C_4$ term.\footnote{
Note that $\mu$ is regarded as a large scale, i.e., $\mu=m_\pi$.
} 
For this reason, the $\tilde{C}_4$ term is
treated as an order higher than the $C_4$ one~\cite{crs-npa99},
and consequently 
the $\tilde{C}_4$ term does not appear (at NNLO) in Eq.~(\ref{eq;aC}).
The other argument is based on the offshell-ness of a term proportional
to $C_4 - \tilde{C}_4$ \cite{fms-npa00}.\footnote{
See the footnote \ref{footnote;C4tildeC4}.
}
In this case, the term proportional to $C_4 - \tilde{C}_4$ is redundant
and thus can be removed by assuming $C_4 = \tilde{C}_4$.
Because both arguments seem to 
have some grounds, to check the dependency of the results
on the values of $C_4$ and $\tilde{C}_4$
we consider the three cases:
1) $\tilde{C}_4=0$ (Ref.~\cite{crs-npa99}),
2) $C_4=\tilde{C}_4$ (Ref.~\cite{fms-npa00}), 
and 3) $C_4=0$. 

In Eq.~(\ref{eq;aC}) there is the $J_0^{div}$ term explicitly
given in Eq.~(\ref{eq;J0div}).  
In the MS scheme used by KR~\cite{kr-plb99,kr-npa00}
one subtracts the infinite term 
$\frac{\alpha M^2}{8\pi}\frac{1}{\epsilon}$
from the $J_0^{div}$. 
One can use another scheme called 
$\overline{\mbox{\rm MS}}$ scheme,
in which finite terms are subtracted together 
with the infinite term so that
$\frac{\alpha M^2}{8\pi}\left[
\frac{1}{\epsilon}-C_E+{\rm ln}(4\pi)
\right]$
is subtracted. 
Then we have 
\bea
J_0^{\overline{MS}} &=&
- \frac{M}{4\pi}\mu
+ \frac{\alpha M^2}{4\pi}\left[
\ln\left(\frac{\mu}{2\alpha M}\right)
+ 1
-C_E 
\right]\, .
\label{eq;J0MSbar}
\eea
This leads to a significant subtraction scheme
dependence in 
the scattering length $a(\mu)$. 

\vskip 3mm \noindent
{\bf 6. Numerical results}

We may define the strong scattering length and the effective range,
respectively, in the zeroth order of $\alpha$ as~\cite{kr-npa00}
\bea
\frac{1}{a(\mu)} = \frac{4\pi}{M}\frac{1}{C_0(\mu)}+\mu\, ,
\ \ \
\frac12 r(\mu) = \frac{4\pi}{M}\frac{C_2(\mu)}{C_0^2(\mu)}\, .
\label{eq;amurmu}
\eea 
Inserting the expressions of $a(\mu)$ and $r(\mu)$ in Eqs.~(\ref{eq;amurmu}) 
into Eqs.~(\ref{eq;aC}) and (\ref{eq;r0}), 
we have 
\bea
\frac{1}{a(\mu)} &=& 
\left[\frac{1}{a(\mu)}\right]_{LO} 
+ \left[\frac{1}{a(\mu)}\right]_{NLO} 
+ \left[\frac{1}{a(\mu)}\right]_{NNLO}\, ,
\label{eq;amu}
\\
r(\mu) &=& 
r_0
- (\alpha M) \left[ D_3 Pr_0^3\mu + D_4 \frac{r_0^2\mu}{\frac{1}{a_C}-\mu}
\right]\, , 
\label{eq;r0mu}
\eea
where 
\begin{eqnarray}
\left[\frac{1}{a(\mu)}\right]_{LO} &=& 
\frac{1}{a_C}
+\alpha M\left[
\ln\left(\frac{\mu}{2\alpha M}\right)+1-C_E \right],
\label{eq;aLO} \\
\left[\frac{1}{a(\mu)}\right]_{NLO} &=& 
-\frac12 \alpha M r_0\mu
- (\alpha M)^2 \left[\frac14 r_0 + \frac{\pi^2}{12}
\frac{r_0 \mu}{\frac{1}{a_C} - \mu} \right],
\label{eq;aNLO} \\ 
\left[\frac{1}{a(\mu)}\right]_{NNLO} &=&
(\alpha M)^2 \left[
D_1 Pr_0^3\mu^2
-\frac{D_2r_0\mu}{12}\frac{r_0\mu}{\frac{1}{a_C}-\mu}
\right] \, ,
\label{eq;aNNLO} 
\end{eqnarray}
and the term linear in $\alpha M$ in Eq.~(\ref{eq;r0mu}) is the 
NNLO correction to $r(\mu)$.
We have three set of coefficients, $X_{x(=1,2,3)}=\{D_1,D_2,D_3,D_4\}$,
because of the additional constraints 
imposed on the LECs $C_4$ and $\tilde{C}_4$ mentioned 
before Eq.~(\ref{eq;J0MSbar}). 
$X_1= \{1,0,4,0\}$ corresponds to the case 1) $\tilde{C}_4=0$,
$X_2= \{7/6,1,10/3,1/6\}$ corresponds to the case 2) $\tilde{C}_4=C_4$, 
and $X_3=\{4/3,-10,8/3,1/3\}$ to the case 3) $C_4=0$. 
We use the values of effective range parameters,
\bea
a_C = - 7.82 \ \mbox{\rm fm}\, , 
\ \ \ 
r_0 = 2.78\ \mbox{\rm fm}\, , 
\ \ \ 
P\simeq 0.022.
\eea
We can also have explicit expressions for the LECs $C_0(\mu)$,
$C_2(\mu)$, $C_4(\mu)$ and $\tilde{C}_4(\mu)$ from Eqs.~(\ref{eq;amu}),
(\ref{eq;r0mu}) and (\ref{eq;P}) with the constraints for 
$C_4$ and $\tilde{C}_4$. 

\begin{figure}[h]
\begin{center}
\epsfig{file=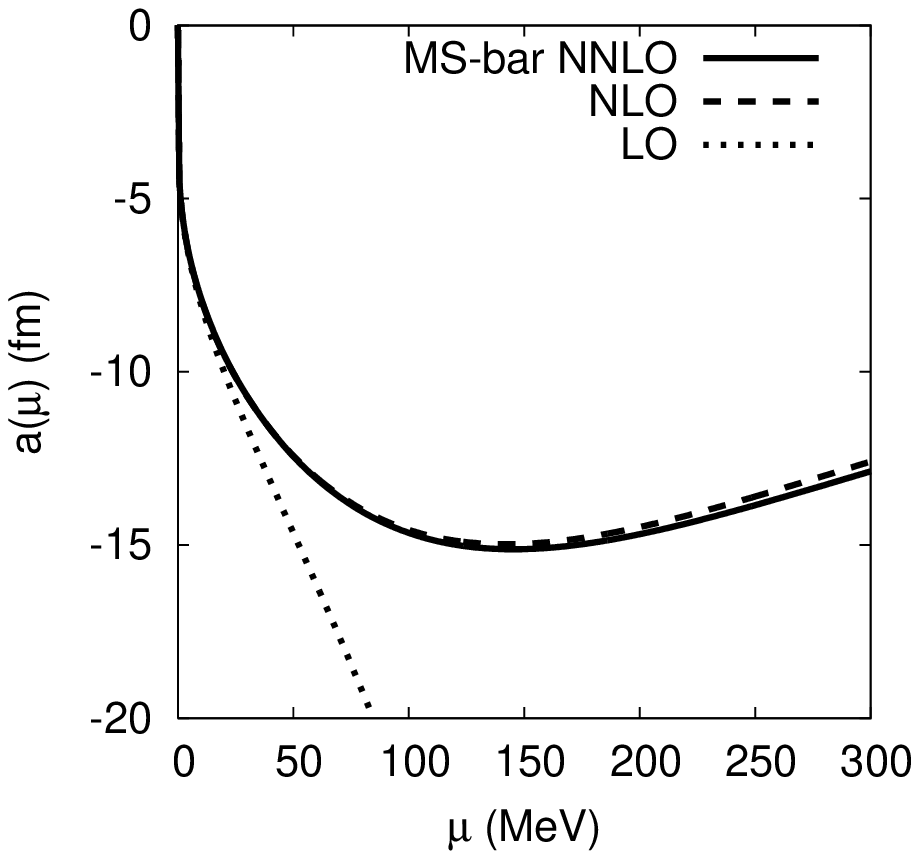, width=7.8cm}
\epsfig{file=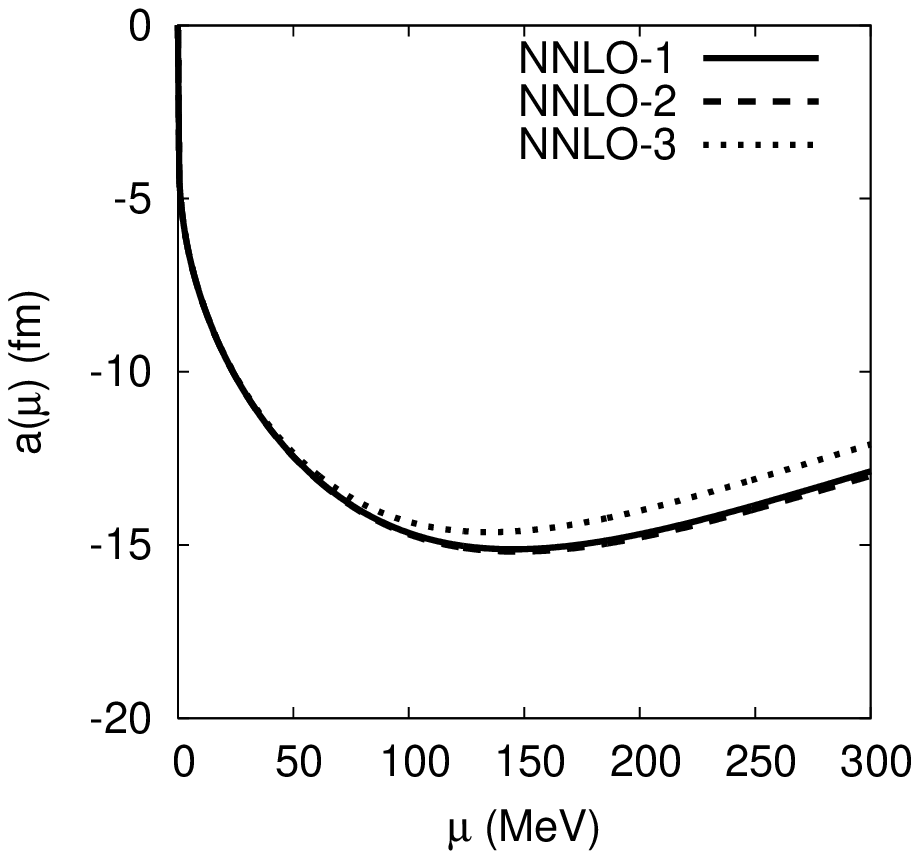, width=7.8cm}
\caption{
Strong scattering length $a(\mu)$ [fm] 
in functions of the scale parameter $\mu$ [MeV].
In the left panel, 
$a(\mu)$ is plotted by the dotted curve,
the dashed curve, and the full curve,
respectively, for up to LO, NLO, and NNLO.
In the right panel, 
$a(\mu)$ calculated up to NNLO
are plotted for 
the three different constraints for $C_4$ and $\tilde{C}_4$,
which are explained in the text.
}
\label{fig;app}
\end{center}
\end{figure}

In Fig.~\ref{fig;app} 
we plot our result of the strong scattering length $a(\mu)$ 
as a function of the scale parameter $\mu$. 
In the left panel, 
we plot three curves for the strong scattering length $a(\mu)$ 
up to LO, NLO, and NNLO with the constraint $\tilde{C}_4=0$ 
(the case 1).
We find that the NLO correction significantly improves the  
estimation of $a(\mu)$, as shown by KR.

If one looks into the details more closely, however, there is 
a quantitative difference 
in the results of LO and NLO between the MS and 
$\overline{\mbox{\rm MS}}$ schemes.
The value of the LO scattering length $a_{LO}(\mu)$
at $\mu=m_\pi$ in the $\overline{\rm MS}$ scheme,
which is obtained from Eq.~(\ref{eq;aLO}),
is $a^{\overline{MS}}_{LO}(\mu=m_\pi) = -30.72$ fm.
The LO contributions 
to $a(\mu)$ can be divided into three terms; $1/a_C$,
the term proportional to a log function 
and the remaining ones proportional to $\alpha M$. 
Evaluating each contribution, we obtain $1/a_C = -0.1279$, 
$\alpha M \ln\left(\frac{m_\pi}{2\alpha M}\right) = 0.0807$
and $\alpha M (1 - C_E) = 0.0147$ in units of ${\rm fm}^{-1}$
in the $\overline{\mbox{\rm MS}}$ scheme.
There is a strong cancellation between $1/a_C$ and the log term
which has the order of $\alpha M$.
Consequently $1/a(\mu=m_\pi)$ becomes a small value, 
making its inverse large.
In the case of the MS scheme, the cancellation is stronger,
having the log term 
$\alpha M \ln\left(\frac{\pi m^2_\pi}{\alpha^2 M^2}\right) = 0.1247$
and
$\alpha M (-3 C_E + 2)/2 = 0.0047$ in units of ${\rm fm}^{-1}$.
The cancellation of $1/a_C$ and the terms proportional to $\alpha M$
makes the value of $1/a(\mu=m_\pi)$ two orders of magnitude smaller 
than $1/a_C$.
As a result, one gets an unrealistically huge scattering length,
$a_{LO}^{MS}(\mu = m_\pi) = 738.62$ fm.
The strong dependence on the renormalization schemes of
the LO contribution to $a(\mu)$ makes the EFT result somehow arbitrary.

The NLO contribution, Eq.~(\ref{eq;aNLO})
can be divided into terms linear in $\alpha M$
and those proportional to $(\alpha M)^2$. 
The term linear in $\alpha M$ is comparable 
in magnitude with the LO contribution 
because of the cancellation in LO, 
as discussed above. 
More precisely, we have 
$1/a^{\overline{MS}}_{LO} = -0.0325$ and 
$-\alpha M r_0 \mu /2 = -0.0343$ in units of ${\rm fm}^{-1}$.
On the other hand, the numerical value of the contribution proportional to 
$(\alpha M)^2$ is 0.0015 in units of ${\rm fm}^{-1}$, 
which is about 5\% of the terms linear in $\alpha M$.

The NNLO contribution is very small,
as can be seen from the left and right panels
in Fig.~\ref{fig;app} and Table~\ref{tab;amu-r0mu}.
The reason can be easily found from the expressions for the NNLO terms
in Eq.~(\ref{eq;aNNLO}).
These terms are proportional to $(\alpha M)^2$.
We observed in NLO that the $(\alpha M)^2$ term is smaller than
the $\alpha M$ order term by an order of magnitude.
The magnitude of $(\alpha M)^2$ terms in NNLO ranges from about 20\%
to 300\% of $(\alpha M)^2$ terms in NLO, depending on the 
choice of the assumptions on $C_4$ and $\tilde{C}_4$.
Consequently, the NNLO correction to $1/a(\mu)$ 
is about $1 \sim 6$\% of the contributions up to NLO, depending on 
the constraints of $C_4$ and $\tilde{C}_4$.

In Table~\ref{tab;amu-r0mu}
we show the estimated values of the strong scattering length $a(\mu)$
and effective range $r(\mu)$ at $\mu=m_\pi$.\footnote{ 
We find a minimum point for $a(\mu)$ at
$\mu\simeq 2/r_0\simeq 142$ MeV,
which is very close to the pion mass, $\mu=m_\pi$.
}
The NNLO term itself varies by an order of magnitude 
depending on the choice of the constraints on $C_4$ and $\tilde{C}_4$. 
However, as discussed in a previous paragraph, its
contribution to $a(\mu)$ is suppressed due to a higher order 
of $\alpha M$ factor.
As a result, the different choice of the constraints 
on $C_4$ and $\tilde{C}_4$ affects little
the final result, only a few percents at most.
The first correction to $r(\mu)$ 
appears at NNLO and  
is linear in $\alpha M$, 
whereas the NLO correction to $1/a(\mu)$ does in the $\alpha M$ order.
Contrary to the case of $1/a(\mu)$ 
where the $\alpha M$ correction plays
a crucial role, the $\alpha M$ contribution to $r(\mu)$ 
amounts to only about 2\% of $r_0$.
Though the $\alpha M$ order corrections to $1/a(\mu)$ and $r(\mu)$
are of the same order of magnitude,
the $(\alpha M)^0$ order contribution to $1/a_C$ is smaller than
that of $r_0$ by an order of magnitude.
Consequently, we have very contrasting behavior of $a(\mu)$ and 
$r(\mu)$.
\begin{table}[tbp]
\begin{center}
\begin{tabular}{c|cccc} \hline
           & NLO& NNLO-1 & NNLO-2 & NNLO-3 \\ \hline
$a(\mu)$   & $-14.98$ & $-15.11$ & $-15.18$ & $-14.62$ \\
$r(\mu)$ & ---     & $2.73$ & 2.78 & 2.82  \\ \hline
\end{tabular}
\caption{
Numerical estimations (in units of fm) 
of scattering length $a(\mu)$ and 
effective range $r(\mu)$ up to NLO and NNLO 
without Coulomb effect at $\mu=140$ MeV.}
\label{tab;amu-r0mu}
\end{center}
\end{table}

Thus our results of the strong $pp$ scattering length
and effective range up to NNLO, which are estimated 
by employing the dimensional 
regularization and the $\overline{\mbox{\rm MS}}$ and PDS schemes
at $\mu=m_\pi$,
can be summarized as
\bea
a(\mu=m_\pi) &=& -14.9 \pm 0.3 \ \ \mbox{\rm fm}\, ,
\\
r(\mu=m_\pi) &=& 2.78 \pm 0.05 \ \ \mbox{\rm fm}\, ,
\eea
where the error-bars are estimated 
by the uncertainties due to the constraints on 
$C_4$ and $\tilde{C}_4$, which could play a similar role to 
the model dependence 
in deducing the values of the strong scattering length $a_{pp}$
and effective range $r_{0,pp}$
in the potential model calculations. 

\vskip 3mm \noindent
{\bf 7. Discussion and conclusions}

In this work, we calculated the $S$-wave $pp$ scattering 
amplitude up to NNLO in the framework of the pionless EFT.
The loop functions were calculated by using the dimensional
regularization with the $\overline{\mbox{\rm MS}}$ and PDS schemes.
After fixing the LECs by using the effective range parameters, we 
estimated the strong scattering length $a(\mu)$ and 
the strong effective range $r(\mu)$ 
as functions of $\mu$.
The LO contributions to $1/a(\mu)$ are composed of
$1/a_C$ and the terms depending on $\alpha M$ 
arising from the loop diagrams.
The smallness of $1/a_C$ makes 
it comparable in magnitude to
the $\alpha M$ terms in the same order. 
Due to the opposite signs
of $1/a_C$ and the $\alpha M$ terms, 
furthermore,
there is a strong cancellation among them
and thus it makes the LO result 
for $1/a(\mu)$ suppressed and sensitive to
the renormalization schemes.
The NLO correction, expanded in powers of $\alpha M$, 
begins with the linear order of $\alpha M$. 
The linear $\alpha M$ order correction to $a(\mu)$ 
is of the same order of magnitude as the $\alpha M$ terms in LO, 
and thus makes the NLO contribution crucial in both of the MS and 
$\overline{\mbox{\rm MS}}$ schemes.
The higher $\alpha M$ order terms in NLO, e.g., 
the terms proportional to $(\alpha M)^2$ are suppressed
to a few percents of the leading contribution, so they can be regarded
as a perturbative corrections to both $a(\mu)$ and $r(\mu)$.
The NNLO terms give us only 
a fairly minor correction to the results up to NLO.
The reason is partly attributed to 
the additional order counting of the NNLO terms
in powers of $\alpha M$: 
The $\alpha M$ order corrections in NNLO begin with $(\alpha M)^2$. 
Similar to the $(\alpha M)^2$ contribution in NLO,
the terms in NNLO 
produces small corrections to the results. 
In conclusion, we can say that our investigation reveals both bright
and shadowy aspects of studying the strong $pp$ scattering length 
in EFT.
Convergence from NLO to NNLO is satisfactory, but the LO and NLO results
are significantly dependent on the renormalization schemes.

Though the quantities of the strong scattering length and effective 
range from the $pp$ scattering 
could be regarded as physical quantities,
it is unlikely that they can be determined unambiguously without
the subtraction scheme and renormalization scale dependence within
the present framework of EFT.
Similar arguments can be found in Refs.~\cite{grs-epjc03,g-epja04}.
Nevertheless, the strong $pp$ scattering length and effective range are 
important ingredients for better understanding of the isospin
nature of the $NN$ interaction. 
The problem of the strong $pp$ scattering length may have to be
approached at various levels, from ``first principle calculations"
like lattice QCD to 
more complex systems in which
$a(\mu)$ (or equivalently $C_0(\mu)$) plays non-trivial roles. 

\vskip 3mm \noindent
{\bf Acknowledgments}

We thank 
Yoonbai Kim for a useful comment on our work.
S.A. thanks F. Ravndal for communications.
S.A. is supported by Korean Research Foundation and 
The Korean Federation of Science and Technology Societies Grant
funded by Korean Government (MOEHRD, Basic Research Promotion Fund):
the Brain Pool program (052-1-6) and KRF-2006-311-C00271.

\vskip 3mm \noindent
{\bf Appendix A: Amplitudes in NNLO}

In this appendix we present expressions of each of the amplitudes
in NNLO in terms of functions, $\psi_{0,2,4}$ and $J_{0,2,22,4}$. 
Detailed calculations of the $\psi$ and $J$ functions are given 
in Appendix B.
From the diagram (a) in Fig.~\ref{fig;nnlo-1},  we have
\bea
T_{SC}^{(4,a)} &=& 
\langle \psi_{\vec{p}'}^{(-)}|\hat{V}_2\hat{G}_C^{(+)}
\sum_{n=0}^\infty(\hat{V}_0\hat{G}_C^{(+)})^n   
\hat{V}_0G_C^{(+)}\hat{V}_2|\psi_{\vec{p}}^{(+)}\rangle \, 
\nnb \\ &=& 
\frac{C_0}{1-C_0J_0}
\int\frac{d^3\vec{q}'}{(2\pi)^3}
\langle \psi_{\vec{p}'}^{(-)}|\hat{V}_2\hat{G}_C^{(+)}|\vec{q}'\rangle
\int\frac{d^3\vec{q}}{(2\pi)^3}
\langle\vec{q}|\hat{G}_C^{(+)}\hat{V}_2|\psi_{\vec{p}}^{(+)}\rangle
\nnb \\ &=&
\frac14 \frac{C_0C_2^2}{1-C_0J_0}
(\psi_0J_2+\psi_2J_0)^2\, .
\eea
From the diagrams (b) and (c) in Fig.~\ref{fig;nnlo-1} we have
\bea
T_{SC}^{(4,b,c)} &=& 
\langle \psi_{\vec{p}'}^{(-)}|
\sum_{n=0}^\infty (\hat{V}_0G_C^{(+)})^n
\hat{V}_0 \hat{G}_C^{(+)}\hat{V}_2\hat{G}_C^{(+)}
\sum_{m=0}^\infty (\hat{V}_0G_C^{(+)})^m
\hat{V}_0G_C^{(+)}\hat{V}_2|\psi_{\vec{p}}^{(+)}\rangle
\nnb \\ && + 
\langle \psi_{\vec{p}'}^{(-)}|
\hat{V}_2\hat{G}_C^{(+)}
\sum_{n=0}^\infty (\hat{V}_0G_C^{(+)})^n
\hat{V}_0 \hat{G}_C^{(+)}\hat{V}_2\hat{G}_C^{(+)}
\sum_{m=0}^\infty (\hat{V}_0G_C^{(+)})^m
\hat{V}_0|\psi_{\vec{p}}^{(+)}\rangle
\nnb \\ &=& 
\frac{C_0^2\psi_0}{(1-C_0J_0)^2}
\int\frac{d^3\vec{q}}{(2\pi)^3}\frac{d^3\vec{q}'}{(2\pi)^3}
 \langle\vec{q}'|\hat{G}_C^{(+)}\hat{V}_2\hat{G}_C^{(+)}|\vec{q}\rangle
\nnb \\ && \times
\int\frac{d^3\vec{k}}{(2\pi)^3}\left[
 \langle \vec{k}|\hat{G}_C^{(+)}\hat{V}_2|\psi_{\vec{p}}^{(+)}\rangle
+ \langle\psi_{\vec{p}'}^{(-)}|\hat{V}_2\hat{G}_C^{(+)}|\vec{k}\rangle
\right]
\nnb \\ &=& 
\frac{C_0^2C_2^2\psi_0J_0J_2}{(1-C_0J_0)^2}(\psi_0J_2+\psi_2J_0)\, .
\eea
From the diagram (d) in Fig.~\ref{fig;nnlo-1}, we have
\bea
T_{SC}^{(4,d)} &=& 
\langle\psi_{\vec{p}'}^{(-)}|
\sum_{l=0}^\infty (\hat{V}_0\hat{G}_C^{(+)})^l \hat{V}_0
\hat{G}_C^{(+)}\hat{V}_2\hat{G}_C^{(+)} 
\sum_{m=0}^\infty (\hat{V}_0\hat{G}_C^{(+)})^m \hat{V}_0
\hat{G}_C^{(+)}\hat{V}_2\hat{G}_C^{(+)} 
\sum_{n=0}^\infty (\hat{V}_0\hat{G}_C^{(+)})^n \hat{V}_0
|\psi_{\vec{p}}^{(+)}\rangle
\nnb \\ &=&
\frac{C_0^3\psi_0^2}{(1-C_0J_0)^3}\left[
\int\frac{d^3\vec{q}'}{(2\pi)^3}\frac{d^3\vec{q}}{(2\pi)^3}
\langle \vec{q}'|\hat{G}_C^{(+)}\hat{V}_2\hat{G}_C^{(+)}|\vec{q}\rangle
\right]^2
= \frac{C_0^3C_2^2\psi_0^2}{(1-C_0J_0)^3}J_0^2J_2^2\, .
\eea
From the diagram (e) in Fig.~\ref{fig;nnlo-2} we have
\bea
T_{SC}^{(4,e)} &=& 
\langle\psi_{\vec{p}'}^{(-)}|
\vec{V}_2\hat{G}_C^{(+)}\hat{V}_2|\psi_{\vec{p}}^{(+)}\rangle
= \frac{C_2^2}{4}(\psi_0^2J_{22}+\psi_2^2J_0+2\psi_0\psi_2J_2)\, .
\eea
From the diagrams (f) and (g) in Fig.~\ref{fig;nnlo-2} we have
\bea
T_{SC}^{(4,f,g)} &=& 
\langle \psi_{\vec{p}'}^{(-)}| 
\sum_{n=0}^\infty (\hat{V}_0\hat{G}_C^{(+)})^n \hat{V}_0
\hat{G}_C^{(+)}\hat{V}_2
\hat{G}_C^{(+)}\hat{V}_2
|\psi_{\vec{p}}^{(+)}\rangle
\nnb \\ && +
\langle \psi_{\vec{p}'}^{(-)}| 
\hat{V}_2 \hat{G}_C^{(+)}
\hat{V}_2 \hat{G}_C^{(+)}
\sum_{n=0}^\infty (\hat{V}_0\hat{G}_C^{(+)})^n \hat{V}_0
|\psi_{\vec{p}}^{(+)}\rangle
\nnb \\ &=& \frac12 \frac{C_0C_2^2\psi_0}{1-C_0J_0}
(\psi_0J_2^2 + 2 \psi_2 J_0 J_2 + \psi_0 J_0 J_{22})\, .
\eea
From the diagram (h) in Fig.~\ref{fig;nnlo-2} we have
\bea
T_{SC}^{(4,h)} &=& 
\langle \psi_{\vec{p}'}^{(-)}|
\sum_{m=0}^\infty (\hat{V}_0\hat{G}_C^{(+)})^m \hat{V}_0
\hat{G}_C^{(+)}\hat{V}_2
\hat{G}_C^{(+)}\hat{V}_2
\hat{G}_C^{(+)}
\sum_{n=0}^\infty (\hat{V}_0\hat{G}_C^{(+)})^n \hat{V}_0
|\psi_{\vec{p}}^{(+)}\rangle
\nnb \\ &=& 
\frac14 \frac{C_0^2C_2^2\psi_0^2}{(1-C_0J_0)^2}(3J_0J_2^2+J_0^2J_{22})\, .
\eea
From the diagram (i) in Fig.~\ref{fig;nnlo-3} we have
\bea
T_{SC}^{(4,i)} &=& 
\langle \psi_{\vec{p}}^{(-)}|\hat{V}_4
|\psi_{\vec{p}}^{(+)}\rangle
=\frac12 C_4\psi_2^2
+ \frac12\tilde{C}_4\psi_0 \psi_4 \, .
\eea 
From the diagrams (j) and (k) in Fig.~\ref{fig;nnlo-3} we have
\bea
T_{SC}^{(4,j,k)} &=& 
\langle \psi_{\vec{p}'}^{(-)}|
\sum_{n=0}^\infty (\hat{V}_0\hat{G}_C^{(+)})^n \hat{V}_0
\hat{G}_C^{(+)}
\hat{V}_4
|\psi_{\vec{p}}^{(+)}\rangle
+ \langle \psi_{\vec{p}'}^{(-)}|
\hat{V}_4
\hat{G}_C^{(+)}
\sum_{n=0}^\infty (\hat{V}_0\hat{G}_C^{(+)})^n \hat{V}_0
|\psi_{\vec{p}}^{(+)}\rangle
\nnb \\ &=& 
\frac12 \frac{C_0\psi_0}{1-C_0J_0}\left[ 2C_4\psi_2J_2
+\tilde{C}_4 (\psi_0J_4+\psi_4J_0) \right] \, .
\eea
From the diagram (l) in Fig.~\ref{fig;nnlo-3} we have
\bea
T_{SC}^{(4,l)} &=&
\langle \psi_{\vec{p}'}^{(-)}|
\sum_{m=0}^\infty (\hat{V}_0\hat{G}_C^{(+)})^m \hat{V}_0
\hat{G}_C^{(+)} \hat{V}_4
\hat{G}_C^{(+)} 
\sum_{n=0}^\infty (\hat{V}_0\hat{G}_C^{(+)})^n \hat{V}_0
|\psi_{\vec{p}}^{(+)}\rangle
\nnb \\ &=& 
\frac12 \frac{C_0^2\psi_0^2}{(1-C_0J_0)^2}\left[
C_4J_2^2 +\tilde{C}_4J_0J_4\right] \, .
\eea

\vskip 2mm \noindent
{\bf Appendix B: Loop functions} 

In this appendix, we 
present $\psi$ functions 
($\psi_0$, $\psi_2$, $\psi_4$) 
and $J$ functions
($J_0$, $J_2$, $J_{22}$, and $J_4$) 
employing dimensional regularization and power divergent regularization 
scheme~\cite{kr-npa00,ksw-plb98}.
We first show the calculations of the $\psi_0$, $\psi_2$, $\psi_4$ functions 
in Eqs.~(\ref{eq;psi0}), (\ref{eq;psi2}), (\ref{eq;psi4}). 

\vskip 1mm \noindent 
{\bf 1.} $\psi_0$:
The Fourier transformation of the Coulomb wavefunction
$\psi^{(\pm)}_{\vec{p}}(\vec{r})$ is 
\bea
\psi^{(\pm)}_{\vec{p}}(\vec{k}) &=&
\int d^3\vec{r}\psi^{(\pm)}_{\vec{p}}(\vec{r})
e^{-i\vec{k}\cdot \vec{r}}\, ,
\eea
with
\bea
\psi_{\vec{p}}^{(\pm)}(\vec{r}) &=& 
\sum_{l=0}^\infty (2l+1) i^l R_l^{(\pm)}(pr) P_l({\rm cos}\,\theta)\, ,
\eea
where ${\rm cos}\theta = \hat{p}\cdot\hat{r}$.
One has the relation,
$\vec{k}\cdot\vec{r}= kr[\cos\theta \cos\hat{\theta}
+\sin\theta\sin\hat{\theta}\cos(\phi-\hat{\phi})]$, 
where $\vec{r}$ and $\vec{k}$ are represented by 
$(r,\theta,\phi)$ and $(k,\hat{\theta},\hat{\phi})$,
respectively. 
Now we choose $\hat{\phi}=0$ and then have 
\bea
\int^{2\pi}_0d\phi e^{-ikr\sin\theta\sin\hat{\theta}\cos\phi}
= 2\pi J_0(-kr\sin\theta\sin\hat{\theta})\, ,
\eea
where $J_n$ is a Bessel function 
and we have used the Bessel's first integral,
$J_n(z) = \frac{1}{2\pi i^n}\int^{2\pi}_0d\phi e^{iz\cos\phi}e^{in\phi}$. 
Using the relations,
\bea
\int^\pi_0d\theta \sin\theta P_l(\cos\theta)
J_0(-kr\sin\theta\sin\hat{\theta})
e^{-ikr\cos\theta\cos\hat{\theta}}
=i^l\sqrt{\frac{2\pi}{-kr}}P_l(\cos\theta)
J_{l+\frac12}(-kr)\, ,
\label{eq;pp}
\eea 
$J_l(-z) = (-1)^l J_l(z)$, 
and 
$j_l(z) = \sqrt{\frac{\pi}{2z}}J_{l+\frac12}(z)$ 
where Eq.~(\ref{eq;pp}) is obtained 
from Eq.~(15) in Ref.~\cite{pp-pr29},
we have
\bea
\psi_{\vec{p}}^{(\pm)}(\vec{k}) &=& 
4\pi \sum_{l=0}^\infty(2l+1) P_l(\cos\hat{\theta})
\int^\infty_0drr^2R_l^{(\pm)}(pr)j_l(kr)\, .
\eea

Now we calculate $\psi_0$ by the dimensional regularization. 
The angular integration will pick up the $l=0$ part
of the wavefunction, thus we have
\bea
\psi_0(p) &=& 
\left(\frac{\mu}{2}\right)^{4-d}
\int\frac{d^{d-1}\vec{k}}{(2\pi)^{d-1}}
\psi_{\vec{p}}^{(+)}(\vec{k})
\nnb \\ &=&
4\pi\left(\frac{\mu}{2}\right)^{4-d}
\frac{\Omega_{d-1}}{(2\pi)^{d-1}}
\int^\infty_0drr^2R_0^{(+)}(pr)
\int^\infty_0dkk^{d-2}j_0(kr)\, 
\nnb \\ &=&
(2\pi)^{3/2}\left(\frac{\mu}{2}\right)^{4-d}
\frac{\Omega_{d-1}}{(2\pi)^{d-1}}
\int^\infty_0drr^{3-d}R_0^{(+)}(pr)
\int^\infty_0d\rho\rho^{d-\frac52}J_{\frac12}(\rho)\,.
\eea
Using the relation
$\int^\infty_0dt\, t^{\alpha-1}J_\nu(t) = 
\frac{2^{\alpha-1}}{\Gamma\left(\frac12(2-\alpha+\nu)\right)}
\Gamma\left(\frac{\alpha+\nu}{2}\right)$, 
we have
\bea
\int^\infty_0d\rho\rho^{d-\frac52}J_\frac12(\rho) &=& 
2^{d-\frac52}
\frac{\Gamma\left(\frac{d-1}{2}\right)}{
\Gamma\left(\frac{4-d}{2}\right)}\, .
\eea
Furthermore, from Eq.~(6.64) of Ref.~\cite{QCT75}
we have
$R_0^{(+)}(pr) = e^{i\sigma_0}C_\eta {}_1F_1(1+i\eta,2;-2ipr)e^{ipr}$, 
where ${}_1F_1(a;b;z)$ is the confluent hypergeometric function
(or Kummer's function of the first kind). 
Using the relation, 
$\int^\infty_0e^{-t}t^{b-1}{}_1F_1(a,c;tz)=\Gamma(b){}_2F_1(a,b,c;z)$,
where ${}_2F_1(a,b;c;z)$ is the first hypergeometric function,
we have
\bea
\int^\infty_0e^{ipr}r^{3-d}{}_1F_1(1+i\eta,2,-2ipr)
=\Gamma(4-d)(-ip)^{d-4}{}_2F_1(1+i\eta,4-d,2;2)\, ,
\eea
and thus
\bea
\psi_0 = (2\pi)^{3/2}
\left(\frac{\mu}{2}\right)^{4-d}
\frac{\Omega_{d-1}}{(2\pi)^{d-1}}
e^{i\sigma_0}C_\eta 
\Gamma(4-d)
(-ip)^{d-4}
{}_2F_1(1+i\eta,4-d,2;2)
2^{d-\frac52}
\frac{\Gamma\left(\frac{d-1}{2}\right)}{
\Gamma\left(\frac{4-d}{2}\right)} \, .
\label{eq;psi0app}
\eea
There are no poles at $d=3$ and 4 in Eq.~(\ref{eq;psi0app}).
Using the relation ${}_2F_1(1+i\eta,0,2;2)=1$ and 
$\Omega_{d}=2\pi^{d/2}/\Gamma(d/2)$, we have
\bea
\psi_0 = e^{i\sigma_0}C_\eta\, .
\eea 

\vskip 1mm \noindent
{\bf 2.} $\psi_2$:
\bea
\lefteqn{\psi_2(p) = \left(\frac{\mu}{2}\right)^{4-d}
\int\frac{d^{d-1}\vec{k}}{(2\pi)^{d-1}}
\psi_{\vec{p}}^{(+)}(\vec{k})\vec{k}^2}
\nnb \\ &=&
(2\pi)^{3/2}
\left(\frac{\mu}{2}\right)^{4-d}
\frac{\Omega_{d-1}}{(2\pi)^{d-1}}
\int^\infty_0drr^{1-d}R_0^{(+)}(pr)
\int^\infty_0d\rho\rho^{d-\frac12}J_\frac12(\rho)
\nnb \\ &=&
(2\pi)^{3/2}
\left(\frac{\mu}{2}\right)^{4-d}
\frac{\Omega_{d-1}}{(2\pi)^{d-1}}
e^{i\sigma_0}C_\eta
(-ip)^{d-2}{}_2F_1(1+i\eta,2-d,2;2)
2^{d-\frac32} \frac{\Gamma\left(\frac{d+1}{2}\right)}{3-d}
\frac{\Gamma(4-d)}{\Gamma\left(\frac{4-d}{2}\right)}\, .
\nnb \\
\eea
For $d=4$ we have
\bea
\psi_2 &=& e^{i\sigma_0}C_\eta\left(p^2 -\frac12\alpha^2 M^2\right)\, ,
\eea
where we have used the relation 
${}_2F_1(1+i\eta,-2,2,2)=\frac13 -\frac23\eta^2$.
For $d=3$ we have 
\bea
\psi^{(d=3)}_2 &=& 
- e^{i\sigma_0}C_\eta\alpha M\mu\frac{1}{3-d}+ \cdots\, ,
\eea
where we have used the relation 
${}_2F_1(1+i\eta,-1,2;2)=-i\eta$, and thus we have 
\bea
\psi_2 &=& e^{i\sigma_0}C_\eta\left[
p^2 
-\alpha M\mu
-\frac12 (\alpha M)^2 
\right] \, .
\label{eq;psi2app}
\eea

\vskip 1mm \noindent
{\bf 3.}  $\psi_4$:
\bea
\lefteqn{\psi_4 =
\left(\frac{\mu}{2}\right)^{4-d}
\int\frac{d^{d-1}\vec{k}}{(2\pi)^{d-1}}
\psi_{\vec{p}}^{(+)}(\vec{k})\vec{k}^4}
\nnb \\ &=&
(2\pi)^{3/2}
\left(\frac{\mu}{2}\right)^{4-d} 
\frac{\Omega_{d-1}}{(2\pi)^{d-1}}
\int^\infty_0drr^{-1-d}R_0^{(+)}(pr)
\int^\infty_0d\rho  \rho^{d+\frac32}J_\frac12(\rho) 
\nnb \\ &=&
(2\pi)^{3/2}
\left(\frac{\mu}{2}\right)^{4-d}
\frac{\Omega_{d-1}}{(2\pi)^{d-1}}
e^{i\sigma_0}
C_\eta (-ip)^d
{}_2F_1(1+i\eta,-d,2;2)
\frac{2^{d+\frac32}\Gamma\left(\frac{d+3}{2}\right)}{4(1-d)(3-d)}
\frac{\Gamma(4-d)}{\Gamma\left(\frac{4-d}{2}\right)}\, .
\nnb \\
\eea
For $d=4$ we have
\bea
\psi_4 &=& 
e^{i\sigma_0}C_\eta 
\left( p^4-\frac56 \alpha^2M^2p^2
+\frac{1}{24}\alpha^4M^4\right)\, ,
\eea
where we have used the relation 
${}_2F_1(1+i\eta,-4,2;2) = \frac{1}{15}(3-10\eta^2+2\eta^4)$.
For $d=3$ we have
\bea
\psi_4^{d=3} 
= -\frac43e^{i\sigma_0}C_\eta \alpha M\mu 
\left(p^2-\frac18 \alpha^2M^2\right) \frac{1}{3-d} + \cdots\, ,
\eea
where we have used the relation 
${}_2F_1(1+i\eta,-3,2;2) = \frac{i}{3}\eta(-2+\eta^2)$. 
Thus we have 
\bea
\psi_4 &=& 
e^{i\sigma_0}C_\eta\left\{
p^4 
-\left[ \frac43 \alpha M\mu + \frac56(\alpha M)^2\right] p^2
+ \frac16 (\alpha M)^3 \mu + \frac{1}{24}(\alpha M)^4 
\right\}
\, .
\label{eq;psi4app}
\eea

\vskip 2mm
Now we calculate loop functions $J_0$, $J_2$, $J_{22}$ and $J_4$
in Eqs.~(\ref{eq;J0}), (\ref{eq;J2}), (\ref{eq;J22}), (\ref{eq;J4}) 
by using the results of the $\psi$ functions obtained above. 

\vskip 2mm \noindent 
{\bf 4. $J_0$}: 
\vskip 1mm
The function $J_0(p)$ is given by~\cite{kr-npa00}
\bea 
J_0(p) = M \int \frac{d^3\vec{l}}{(2\pi)^3}
\frac{2\pi\eta(l)}{e^{2\pi\eta(l)}-1}
\frac{1}{p^2-l^2+i\epsilon}\, ,
\eea
where $l=|\vec{l}|$. 
We now separate $J_0$ into two parts as~\cite{kr-npa00}
\bea
J_0(p) = J_0^{div} + J_0^{fin}\, ,
\eea
where
\bea
J_0^{div} &=& 
- M \int\frac{d^3\vec{l}}{(2\pi)^3} 
\frac{2\pi\eta(l)}{e^{2\pi\eta(l)}-1} \frac{1}{l^2}\, ,
\\
J_0^{fin} &=&
M \int\frac{d^3\vec{l}}{(2\pi)^3} 
\frac{2\pi\eta(l)}{e^{2\pi\eta(l)}-1} \frac{1}{l^2}
\frac{p^2}{p^2-l^2+i\epsilon} \, .
\eea

As $J_0^{fin}$ is already calculated in Ref.~\cite{kr-npa00},
by changing the parameter $x=2\pi\eta(l)$ and using the relation
\bea \int^\infty_0dx \frac{x}{(e^x-1)(x^2+a^2)}
= \frac12\left[
{\rm ln}\left(\frac{1}{2\pi}\right) 
-\frac{\pi}{a}-\psi\left(\frac{1}{2\pi}\right)
\right]\, ,
\eea
where $\psi$ is the logarithmic derivative of the $\Gamma$-function,
we have
\bea
J_0^{fin} &=& - \frac{\alpha M^2}{4\pi}H(\eta)
= -\frac{\alpha M^2}{4\pi}h(\eta) - C_\eta^2 \frac{M}{4\pi}(ip)\, ,  
\eea
where $\eta=\alpha M/(2p)$, 
$H(\eta) = \psi(i\eta) + \frac{1}{2i\eta} -{\rm ln}(i\eta)$, 
and $h(\eta)=Re H(\eta)$. 

Next we calculate the divergence part $J_0^{div}$
in $d=4-2\epsilon$ dimension
\bea
J_0^{div} &=& 
-M\left(\frac{\mu}{2}\right)^{4-d}
\int \frac{d^{d-1}\vec{q}}{(2\pi)^{d-1}}
\frac{2\pi\eta(q)}{e^{2\pi\eta(q)}-1}\frac{1}{q^2} \, .
\eea
Changing the variable $x=2\pi\eta(q)=\pi\alpha M/q$, we have
\bea
J_0^{div} &=& 
-M\left(\frac{\mu}{2}\right)^{4-d}
\frac{2\pi^{(d-1)/2}}{(2\pi)^{d-1}\Gamma\left(\frac{d-1}{2}\right)}
(\alpha \pi M)^{d-3}\int^\infty_0dx \frac{x^{3-d}}{e^x-1} 
\nnb \\ &=&
-M\left(\frac{\mu}{2}\right)^{4-d}
\frac{2\pi^{(d-1)/2}}{(2\pi)^{d-1}\Gamma\left(\frac{d-1}{2}\right)}
(\alpha \pi M)^{d-3}\Gamma(4-d)\zeta(4-d)\, ,
\eea
where we have used the relation $\Omega_d=2\pi^{d/2}/\Gamma(d/2)$
and $\zeta(z)$ is the Riemann's zeta function.
For $d=4-2\epsilon$ we have
\bea
J_0^{div} &=& 
\frac{\alpha M^2}{8\pi}\left[
\frac{1}{\epsilon}-3\gamma +2
+{\rm ln}\left(\frac{\pi\mu^2}{\alpha^2M^2}\right)
\right]\, .
\eea
We also consider the pole for $d=3$, known
as the power divergence subtraction (PDS) scheme pole.
Using the relation
$\lim_{s\to 1}\left[\zeta(s)-\frac{1}{s-1}\right]=\gamma$, 
we have the pole at 3-dimension
\bea
J^{div}_0 &=& -\frac{\mu M}{4\pi}\frac{1}{3-d}+\cdots\, ,
\eea
and thus we include the PDS counter term and have
\bea
J_0^{div} &=&  - \frac{M}{4\pi}\mu 
+ \frac{\alpha M^2}{8\pi}\left[
\frac{1}{\epsilon}-3\gamma +2
+{\rm ln}\left(\frac{\pi\mu^2}{\alpha^2M^2}\right)
\right]\, .
\eea

\vskip 2mm \noindent
{\bf 5.} $J_2$
\bea
J_2 &=& 
\int\frac{d^3\vec{q}}{(2\pi)^3}\frac{d^3\vec{q}'}{(2\pi)^3}
\vec{q}'^2\langle \vec{q}'|\hat{G}_C^{(+)}|\vec{q}\rangle \,
= M\int\frac{d^3\vec{q}}{(2\pi)^3}
\frac{ \psi_2(q)\psi_0^*(q)}
 {\vec{p}^2-\vec{q}^2+i\epsilon} \, .
\eea
Using the result of $\psi_2$ in Eq.~(\ref{eq;psi2app}),
we get 
\bea
J_2(p) &=& \left[
p^2 -\mu\alpha M-\frac12 (\alpha M)^2\right] J_0(p)
-\Delta J_2\, ,
\eea 
where
\bea
\Delta J_2 = M\left(\frac{\mu}{2}\right)^{4-d}
\!\!\!\int\frac{d^{d-1}\vec{k}}{(2\pi)^{d-1}} \frac{2\pi \eta(k)}{
e^{2\pi\eta(k)}-1}
= M\left(\frac{\mu}{2}\right)^{4-d}\!\!\!\!
\frac{\Omega_{d-1}}{(2\pi)^{d-1}}
(\pi \alpha M)^{d-1}\Gamma(2-d)
\zeta(2-d) .
\eea
For $d=4$ we have
\bea
\Delta J_2 &=& \frac14\pi \alpha^3 M^4\zeta'(-2)\, ,
\eea
where $\zeta'(-2)=-0.0304\cdots$.
For $d=3$ we obtain
\bea
\Delta J_2^{(d=3)} &=& \frac{1}{48}\pi\alpha^2 M^3 \mu\frac{1}{3-d} 
+ \cdots,
\eea 
and by including the PDS counter term we have
\bea
\Delta J_2 &=&  
\frac{\pi M}{48}(\alpha M)^2\mu
+ \frac{\pi M}{4}(\alpha M)^3\zeta'(-2) \,.
\eea
 
\vskip 2mm \noindent
{\bf 6.} $J_{22}$
\bea
J_{22} &=& 
\int\frac{d^3\vec{q}}{(2\pi)^3}\frac{d^3\vec{q}'}{(2\pi)^3}
\vec{q}'^2\langle \vec{q}'|\hat{G}_C^{(+)}|\vec{q}\rangle\vec{q}^2 \,
= M \int \frac{d^3\vec{q}}{(2\pi)^3}
\frac{\psi_2(q)\psi_2^*(q)}{p^2-q^2+i\epsilon}
\nnb \\ &=&
(p^4-2Ap^2+A^2)J_0 
-(p^2-2A)\Delta J_2 
-\Delta J_{22}\, ,
\eea
where
$A = \mu\alpha M+\frac12 (\alpha M)^2$, 
and 
\bea
\Delta J_{22} &=& 
M\left(\frac{\mu}{2}\right)^{4-d} 
\int\frac{d^{d-1}\vec{q}}{(2\pi)^{d-1}}\vec{q}^2 
\psi_0(q)\psi_0^*(q)\, 
\nnb \\ &=& M
\left(\frac{\mu}{2}\right)^{4-d}
\frac{\Omega_{d-1}}{(2\pi)^{d-1}}
(\pi \alpha M)^{d+1}\Gamma(-d)\zeta(-d)\, .
\eea
For $d=4$ we have
\bea
\Delta J_{22} &=& \frac{1}{48}\pi^3\alpha^5 M^6 \zeta'(-4)\, ,
\eea
where $\zeta'(-4) = 0.00798\cdots$.
For $d=3$ we get 
\bea
\Delta J_{22}^{(d=3)} &=& - \frac{1}{2880}\pi^3\alpha^4M^5
\mu \frac{1}{3-d}+\cdots\, ,
\eea
and thus we obtain 
\bea
\Delta J_{22} &=& -\frac{\pi^3 M}{2880}(\alpha M)^4\mu
+ \frac{\pi^3M}{48}(\alpha M)^5\zeta'(-4)\, .
\eea

\vskip 2mm \noindent
{\bf 7.} $J_4$
\bea
J_4 &=&
\int\frac{d^3\vec{q}}{(2\pi)^3}\frac{d^3\vec{q}'}{(2\pi)^3}
\vec{q}'^4\langle \vec{q}'|\hat{G}_C^{(+)}|\vec{q}\rangle \,
= M\int\frac{d^3\vec{q}}{(2\pi)^3}\frac{\psi_4(q)\psi_0^*(q)}{
p^2-q^2+i\epsilon}\, .
\eea
Using the relation for $\psi_4$ in Eq.~(\ref{eq;psi4app}),
we have
\bea
J_4 &=& \left\{p^4
-
\left[\frac43 \alpha M\mu +\frac56(\alpha M)^2
\right] p^2 
+\frac16(\alpha M)^3 \mu +\frac{1}{24}(\alpha M)^4
\right\} J_0
\nnb \\ &&
-\left[p^2
- \frac43 \alpha M\mu -\frac56(\alpha M)^2
\right]\Delta J_2
-\Delta J_{22}\, .
\eea

\end{document}